\documentclass[10pt,twocolumn,a4paper]{IEEEtran}
\usepackage{amsmath}
\usepackage{graphicx}

\newcommand{\figuramedia}[3]
{
\begin{figure}
  \centering
 \includegraphics[width=5.5cm]{#1}
  \caption{#2}\label{#3}
\end{figure}
}

\newcommand{\figura}[3]
{
\begin{figure}
  \centering
 \includegraphics[width=8cm]{#1}
  \caption{#2}\label{#3}
\end{figure}
}
\title{A Model of the IEEE 802.11 DCF in Presence of Non Ideal Transmission
Channel and Capture Effects}
\author{\authorblockN{F. Daneshgaran, M. Laddomada, F. Mesiti, and M.
Mondin
\thanks{\textbf{Accepted to IEEE Globecom 2007}}
\thanks{This work has been supported by PRIN-2005 ICONA.}
\thanks{F. Daneshgaran is with ECE Dept., California State University,
Los Angeles, USA.}
\thanks{M. Laddomada (\textrm{laddomada@polito.it}), F. Mesiti and M. Mondin
are with DELEN, Politecnico di Torino, Italy.}} }
\begin{document}
\maketitle

\begin{abstract}
In this paper, we provide a throughput analysis of the IEEE 802.11
protocol at the data link layer in non-saturated traffic
conditions taking into account the impact of both transmission channel
and capture effects in Rayleigh fading environment. Impacts of
both non-ideal channel and capture become important in terms of the
actual observed throughput in typical network conditions whereby
traffic is mainly unsaturated, specially in an
environment of high interference.

We extend the multi-dimensional Markovian state transition model
characterizing the behavior at the MAC layer by including
transmission states that account for packet transmission failures
due to errors caused by propagation through the channel, along
with a state characterizing the system when there are no packets
to be transmitted in the buffer of a station.
\end{abstract}

\section{Introduction}
Wireless Local Area Networks (WLANs) using the IEEE802.11 series
of standards have experienced an exponential growth in the recent
past~\cite{standard_DCF_MAC}-\cite{hamilton}.
The Medium Access Control (MAC) layer of many wireless protocols resemble
that of IEEE802.11. Hence, while we focus on this protocol, it is evident
that the results easily extend to other protocols with similar MAC
layer operation.
%
%

The most relevant works to what is presented here are
\cite{Bianchi,Liaw}. In~\cite{Bianchi} the author provided an
analysis of the saturation throughput of the basic 802.11 protocol
assuming a two dimensional Markov model at the MAC layer, while
in~\cite{Liaw} the authors extended the underlying model in order
to consider unsaturated traffic conditions by introducing a new
idle state, not present in the original Bianchi's model, accounting for
the case in which the station buffer is empty, after a successful completion of a packet
transmission. In the modified model,
however, a packet is discarded after $m$ backoff stages, while in
the Bianchi's model, the station keeps iterating in the $m$-th
backoff stage until the packet gets successfully transmitted.

In \cite{Qiao}, the authors look at the impact of channel induced
errors and the received SNR on the achievable throughput in a
system with rate adaptation whereby the transmission rate of the
terminal is adapted based on either direct or indirect
measurements of the link quality. In \cite{Chatzimisios}, the
authors deal with the extension of Bianchi's Markov model in order
to account for channel errors. Paper~\cite{Malone} proposes an
extension of the Bianchi's model considering a new state for each
backoff stage accounting for the absence of new packets to be
transmitted, i.e., in unloaded traffic conditions.

In real networks, traffic is mostly unsaturated, so it is
important to derive a model accounting for practical network
operations. In this paper~\cite{Daneshgaran}, we extend the
previous works on the subject by looking at all the three issues
outlined before together, namely real channel conditions,
unsaturated traffic, and capture effects. Our assumptions are
essentially similar to those of Bianchi~\cite{Bianchi} with the
exception that we do assume the presence of both channel induced
errors and capture effects due to the transmission over Rayleigh
fading channel.

The paper is organized as follows. Section~II extends the Markov
model initially proposed by Bianchi, presenting modifications that
account for transmission errors and capture effects over Rayleigh
fading channels employing the 2-way handshaking technique in
unsaturated traffic conditions. Section~III provides an expression
for the unsaturated throughput of the link. In section~IV we
present simulation results where typical MAC layer parameters for
IEEE802.11b are used to obtain throughput values as a function of
various system level parameters, capture probability, and SNR
under typical traffic conditions. Finally, Section~V is devoted to
conclusions.
\section{Development of the Markov Model}
In \cite{Bianchi}, an analytical model is presented for the
computation of the throughput of a WLAN using the IEEE 802.11 DCF
under ideal channel conditions. By virtue of the strategy employed
for reducing the collision probability of the packets transmitted
from the stations attempting to access the channel simultaneously,
a random process $b(t)$ is used to represent the backoff counter
of a given station. Backoff counter is decremented at the start of
every idle backoff slot and when it reaches zero, the station
transmits and a new value for $b(t)$ is set.

The value of $b(t)$ after each transmission depends on the size of
the contention window from which it is drawn. Therefore it depends
on the station's transmission history, rendering it a
non-Markovian process. To overcome this problem and get to the
definition of a Markovian process, a second process $s(t)$ is defined
representing the size of the contention window
from which $b(t)$ is drawn,  $(W_i=2^i W,~i=s(t))$.

A two-dimensional Markov process $(s(t),b(t))$ can now be defined,
based on two assertions:
\begin{enumerate}
\item the probability $\tau$ that a station will attempt
transmission in a generic time slot is constant across all time
slots;
\item the probability $P_{col}$ that any transmission experiences
a collision is constant and independent of the number of
collisions already suffered.
\end{enumerate}

The main aim of this section is to propose an effective
modification of the bi-dimensional Markov process proposed
in~\cite{Bianchi} in order to account for unsaturated traffic
conditions, channel error propagation and capture effects over a
Rayleigh fading channel under the hypothesis of employing a 2-way
handshaking access mechanism.

On the basis of this assumption, collisions can occur with
probability $P_{col}$ on the transmitted packets, while
transmission errors due to the channel, can occur with probability
$P_e$. We assume that collisions and transmission error events are
statistically independent. In this scenario, a packet is
successfully transmitted if there is no collision (this event has
probability $1-P_{col}$) and the packet encounters no channel
errors during transmission (this event has probability $1-P_{e}$).
The probability of successful transmission is therefore equal to
$(1-P_{e})(1-P_{col})$, from which we can set an equivalent
probability of failed transmission as $P_{eq}=P_e+P_{col}-P_e
P_{col}$.

Furthermore, in mobile radio environment, it may happen that the
channel is captured by a station whose power level is stronger
than other stations transmitting at the same time. This may be
due to relative distances and/or channel conditions for each user
and may happen whether or not the terminals exercise power
control. Capture effect often reduces the
collision probability on the channel since the stations whose
power level at the receiver are very low due to path attenuation,
shadowing and fading, are considered as interferers at the AP
raising the noise floor.
\figuramedia{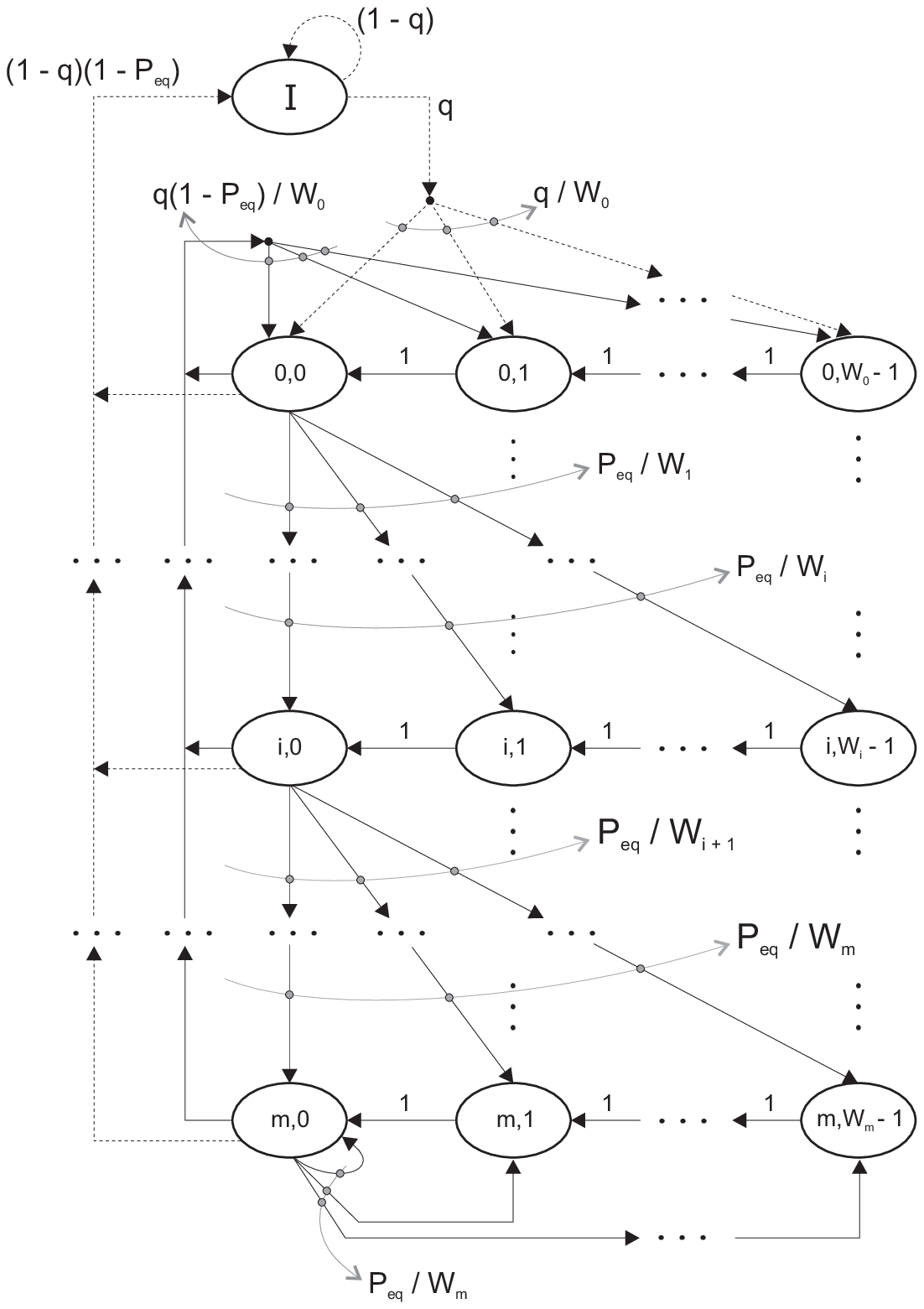}{Markov chain for the contention
model in unsaturated traffic conditions, based on the 2-way
handshaking technique, considering the effects of capture and
channel induced errors.}{fig.chain}

To simplify the analysis, we make the assumption that the impact
of the channel induced errors on the RTS, CTS and Acknowledgment
(ACK) packets are negligible because of their short length. This
is justified on the basis of the assumption that the bit errors
inflicting the transmitted data are independent of each other.
%
%
%
We note that with sufficient interleaving we can always ensure
that the errors inflicting individual bits in a data packet are
independent of each other \cite{laddomada_3,laddomada_2}. 

Practical networks operate in unsaturated traffic conditions. In
this case, Bianchi's model~\cite{Bianchi} assuming the presence of
a packet to be transmitted in each and every station's buffer, is
not valid anymore. However, the simplicity of such a model can be
retained also in unsaturated conditions by introducing a new
state, labelled $I$, accounting for the following two situations:
\begin{itemize}
    \item immediately after a successful transmission, the buffer of the
    transmitting station is empty;
    \item the station is in an idle state with an empty buffer until a new
    packet arrives at the buffer for transmission.
\end{itemize}

With these considerations in mind, let us discuss the Markov model
shown in Fig.~\ref{fig.chain}, modelling unsaturated traffic
condition.
The Markov Process of Fig.~\ref{fig.chain} is governed by the
following transition probabilities\footnote{$P_{i,k|j,n}$ is short
for $P\{s(t+1)=i,b(t+1)=k|s(t)=j,b(t)=n\}$.}:
\begin{equation}\label{eq.process}\small
\begin{array}{lll}
P_{i,k|i,k+1} &= 1,                     &~ k \in [0,W_i-2], ~ i
\in [0,m] \\
P_{0,k|i,0}  &= q(1-P_{eq})/W_0, &~ k \in [0,W_0-1], ~ i \in [0,m]
\\ P_{i,k|i-1,0}   &= P_{eq}/W_i,&~ k \in [0,W_i-1], ~ i \in [1,m]
\\ P_{m,k|m,0}   &= P_{eq}/W_m,&~ k \in [0,W_m-1]\\
P_{I|i,0}  &= (1-q)(1-P_{eq}), &~ i \in [0,m]\\
P_{0,k|I}  &= q/W_0, &~ k \in [0,W_0-1] \\
P_{I|I}  &= 1-q &
\end{array}
\end{equation}
The first equation in~(\ref{eq.process}) states that, at the
beginning of each slot time, the backoff time is decremented. The
second equation accounts for the fact that after a successful
transmission, a new packet transmission starts with backoff stage
0 with probability $q$, in case there is a new packet in the buffer to be transmitted .
Third and fourth equations deal with
unsuccessful transmissions and the need to reschedule a new
contention stage. The fifth equation deals with the practical
situation in which after a successful transmission, the buffer of
the station is empty, and as a consequence, the station transits
in the idle state $I$ waiting for a new packet arrival.
The sixth equation models the situation in which a new packet
arrives in the station buffer, and a new backoff procedure is
scheduled. Finally, the seventh equation models the situation in
which there are no packets to be transmitted and the station is in
the idle state.
\section{Markovian Process Analysis and Throughput Computation}
Next line of pursuit consists in finding a solution of
the stationary distribution:
\[
b_{i,k}=\lim_{t\rightarrow \infty}P[s(t)=i,b(t)=k],~\forall
k\in[0,W_i-1],~\forall i\in[0,m]
\]
that is, the probability of a station occupying a given state at
any discrete time slot. First, we note the following relations:
\begin{equation}\label{trans_states_probabilities}
\begin{array}{rcll}
b_{i,0} & = & P_{eq} \cdot b_{i-1,0} = P_{eq}^i \cdot b_{0,0},   &\forall i \in [1,m-1] \\
b_{m,0} & = & \frac{P_{eq}^m}{1-P_{eq}} \cdot b_{0,0},           & i = m \\
\end{array}
\end{equation}
whereby $P_{eq}$ is the equivalent probability of failed
transmission, that takes into account the need for a new
contention due to either packet collision ($P_{col}$) or channel
errors ($P_e$), i.e., $P_{eq}=P_{col}+P_e-P_e\cdot  P_{col}$.

State $b_I$ in Fig.~\ref{fig.chain} considers both the situation
in which after a successful transmission there are no packets to
be transmitted, and the situation in which the packet queue is
empty and the station is waiting for new packet arrival.
The stationary probability to be in state $b_I$ can
be evaluated as follows:
\begin{equation}
\label{eq:b_N}
\begin{array}{rcl}
b_I & = & (1-q)(1-P_{eq}) \cdot \sum_{i=0}^{m}b_{i,0} + (1-q) \cdot b_I \\
    & = & \frac{(1-q)(1-P_{eq})}{q} \cdot \sum_{i=0}^{m}b_{i,0} \\
\end{array}
\end{equation}
The expression above reflects the fact that state $b_I$ can be reached after a successful packet
transmission from any state $b_{i,0},~\forall i \in [0,m]$ with
probability $(1-q)(1-P_{eq})$, or because the station is waiting
in idle state with probability $(1-q)$,
whereby $q$ is the probability of having at least one packet to be
transmitted in the buffer. The statistical model of $q$ will be
discussed in the next section.

The other stationary probabilities for any $k\in[1,W_i-1]$ follow
by resorting to the state transition diagram shown in
Fig.~\ref{fig.chain}:
\begin{equation}\label{eq.bik}
b_{i,k} = \frac{W_i-k}{W_i} \left\{
\begin{array}{ll}
q(1-P_{eq}) \cdot \sum_{i=0}^{m}b_{i,0} + q \cdot b_I,       & i = 0 \\
P_{eq} \cdot b_{i-1,0},                                      & i \in [1,m-1] \\
P_{eq} (b_{m-1,0} + b_{m,0}),                                & i = m \\
\end{array}\right.
\end{equation}
%
%

Employing the normalization condition, after lengthy algebra, and
remembering the relation $ \sum_{i=0}^{m}b_{i,0} =
\frac{b_{0,0}}{1-P_{eq}} $, it is possible to obtain:
\begin{equation}\small
%
1 = \frac{b_{0,0}}{2} \left[ W\left( \sum_{i=0}^{m-1}(2P_{eq})^i +
\frac{(2P_{eq})^m}{1-P_{eq}} \right) + \frac{1}{1-P_{eq}}
        + \frac{2(1-q)}{q} \right]
\end{equation}

Normalization condition yields the following equation for
computation of $b_{0,0}$:
\begin{equation}\label{eqb_00_norm} 
\begin{array}{ll}
 \frac{2(1-P_{eq})(1-2P_{eq})q}
                   {q[(W+1)(1-2P_{eq}) + WP_{eq}(1-(2P_{eq})^m)] +
                   2(1-q)(1-P_{eq})(1-2P_{eq})}
                   \end{array}
\end{equation}

Equ.~(\ref{eqb_00_norm}) is then used to compute $\tau$, the
probability that a station starts a transmission in a randomly
chosen time slot:
\begin{equation}\label{eq.tau}
\begin{array}{ll}
      \tau= \frac{2(1-2P_{eq})q}{q[(W+1)(1-2P_{eq}) + WP_{eq}(1-(2P_{eq})^m)] + 2(1-q)(1-P_{eq})(1-2P_{eq})}
\end{array}
\end{equation}

The collision probability needed to compute $\tau$ can be found
considering that using a 2-way hand-shaking mechanism, a packet from a
transmitting station encounters a collision if in a given time slot, at
least one of the remaining $(N-1)$ stations transmits
simultaneously another packet, and there is no capture. In our
model, we assume that capture is a subset of the collision events.
This is indeed justified by the fact that there is no capture
without collision, and that capture occurs only because of
collisions between a certain number of transmitting stations
attempting to transmit simultaneously on the channel.
\begin{equation}\label{eq.col}
P_{col} = 1-(1-\tau)^{N-1}-P_{cap}
\end{equation}

As far as the capture effects are concerned, we resort to the
mathematical formulation proposed in \cite{zorzi_rao,Spasenovski}.
In particular, under the hypothesis of power-controlled stations
in infrastructure mode, the capture probability conditioned on $i$
interfering frames can be defined as follows:
\begin{equation}\label{eq.capture_conditional}
P_{cp}\left(\gamma>z_o
g(S_f)|i\right)=\frac{1}{{\left[1+z_{0}g(S_{f})\right]}^{i}}
\end{equation}
whereby, $\gamma$, defined as $P_u/\sum_{k=1}^{i}P_k$, is the ratio
of the power $P_u$ of the useful signal and the sum of the
powers of the $i$ interfering channel contenders transmitting
simultaneously $i$ frames, $g(S_f)$ is the inverse of the processing gain of the
correlation receiver, and $z_0$ is the capture ratio, i.e., the
value of the signal-to-interference power ratio identifying the
capture threshold at the receiver. Notice that
(\ref{eq.capture_conditional}) signifies the fact that capture
probability corresponds to the probability that the power ratio
$\gamma$ is above the capture threshold $z_{0}g(S_{f})$ which
considers the inverse of the processing gain $g(S_{f})$. For Direct Sequence
Spread Spectrum (DSSS) using a 11-chip spreading factor
($S_f=11$), we have $g(S_{f})=\frac{2}{3S_f}$.

Upon defining the probability of generating exactly $i+1$
interfering frames over $N$ contending stations in a generic slot
time:
\[
{N \choose i+1}\tau^{i+1}(1-\tau)^{N-i-1}
\]
the frame capture probability $P_{cap}$ can be obtained as follows:
\begin{equation}\label{capture_probability}
P_{cap}=\sum_{i=1}^{N-1}{N \choose
i+1}\tau^{i+1}(1-\tau)^{N-i-1}P_{cp}\left(\gamma>z_o
g(S_f)|i\right)
\end{equation}

Putting together Equ.s~(\ref{eq.tau}),~(\ref{eq.col}),
and~(\ref{capture_probability}), along with $P_{eq}$, the
following nonlinear system can be defined and solved numerically,
obtaining the values of $\tau$, $P_{col}$, $P_{cap}$, and
$P_{eq}$:
\begin{equation}\label{eq.system}
\left\{ \begin{array}{ll}
\tau  =\frac{2(1-2P_{eq})q}{q[(W+1)(1-2P_{eq}) + WP_{eq}(1-(2P_{eq})^m)] + 2(1-q)(1-P_{eq})(1-2P_{eq})}\\
P_{col} = 1-(1-\tau)^{N-1} -P_{cap}\\
P_{eq}   =
P_{col}+P_e-P_e\cdot  P_{col}\\
P_{cap}=\sum_{i=1}^{N-1}{N \choose
i+1}\tau^{i+1}(1-\tau)^{N-i-1}\frac{1}{{(1+z_{0}g(S_{f}))}^{i}}
\end{array} \right.
\end{equation}
The final step in the analysis is the computation of the
normalized system throughput, defined as the fraction of time the
channel is used to successfully transmit payload bits:
\begin{equation}\small
\label{eq.system2} S = \frac{P_t \cdot P_s\cdot
(1-P_e)E[PL]}{(1-P_t
)\sigma+P_t(1-P_s)T_c+P_tP_s(1-P_e)T_s+P_tP_sP_eT_e}
\end{equation}
where the meaning of the underlined symbols is as follows.

$P_t$ is the probability that there is at least one transmission
in the considered time slot, with $N$ stations contending for the
channel, each transmitting with probability $\tau$:
\begin{equation}
\label{equat_Pt} P_t=1-(1-\tau)^N
\end{equation}

$P_s$ is the conditional probability that a packet transmission
occurring on the channel is successful. This event corresponds to
the case in which exactly one station transmits in a given time
slot, or two or more stations transmit simultaneously and capture
by the desired station occurs. These conditions yields the
following probability:
\begin{equation}
\label{equat_PS}
P_s=\frac{N\tau(1-\tau)^{N-1}+P_{cap}}{P_t}
\end{equation}

$T_c$, $T_e$ and $T_s$ are the average times a channel is sensed
busy due to a collision, error affected data frame transmission
time and successful data frame transmission times, respectively.
Knowing the time durations for ACK frames, ACK timeout, DIFS,
SIFS, $\sigma$, data packet length ($PL$) and PHY and MAC headers
duration ($H$), and propagation delay $\tau_p$, $T_c$, $T_s$, and
$T_e$ can be computed as suggested in \cite{kong}. $E[PL]$ is the
average packet payload length. $\sigma$ is the duration of an
empty time slot.
\subsection{Modelling offered load and estimation of probability $q$}
In our analysis, the offered load related to each station is
characterized by parameter $\lambda$ representing the rate at
which packets arrive at the station buffer from the upper layers,
and measured in $pkt/s$. The time between two packet arrivals is
defined as \textit{interarrival time}, and its mean value is
evaluated as $\frac{1}{\lambda}$. One of the most commonly used traffic
models assumes packet arrival process is Poisson. The resulting interarrival
times are exponentially distributed.

In the proposed model, we need a probability, identified by $q$,
that indicates if at the end of a given transmission there is at least
one packet in the queue to be transmitted. Probability $q$ can be
well approximated in a situation with small buffer size \cite{hamilton} through the
following relation:
\begin{equation}
\label{eq:q_prob} q = 1 - e^{- \lambda E[S_{ts}]}
\end{equation}
where, $E[S_{ts}]$ is the \textit{expected time per slot}, which is
useful to relate the state of the Markov chain with the actual
time spent in each state.

A more accurate model can be derived upon considering different
values of $q$ for each backoff state. However, a
reasonable solution consists in using a mean probability valid for
the whole Markov model~\cite{hamilton}, derived from $E[S_{ts}]$.
The value of $E[S_{ts}]$ can be obtained by resorting to the
durations and the respective probabilities of the idle slot
($\sigma$), the successful transmission slot ($T_s$), the error
slot due to collision ($T_c$), and the error slot due to channel
($T_e$), as follows:
\begin{equation}
\label{equ_E_Sts}
\begin{array}{rcl}
E[S_{ts}] & = & (1-P_t) \cdot \sigma + P_t(1-P_s) \cdot T_c +\\
     &   &  +P_t P_s P_e \cdot T_e + P_t P_s (1-P_e) \cdot T_s
\end{array}
\end{equation}
Upon recalling that packet inter-arrival times are exponentially distributed,
we can use the average slot time to calculate the probability $q$
that in such a time interval a given station receives a packet
from upper layers in its transmission queue. The probability that
in a generic time $T$, $k$ events occur, is:
\begin{equation}
P\{a(T) = k\} = e^{- \lambda T} \frac{(\lambda T)^k}{k!}
\end{equation}
from which we obtain the relation (\ref{eq:q_prob}) referred to earlier:
\begin{equation}
\label{eq:q_prob2} q = 1 - P\{a(E[S_{ts}]) = 0\} = 1 - e^{-
\lambda E[S_{ts}]}
\end{equation}
%
%
\section{Simulation Results and Model Validations}
\label{Simulation_results_section}
\begin{table}\caption{Typical network parameters}
\begin{center}
\begin{tabular}{c|c||c|c}\hline
\hline MAC header & 24 bytes  & $\tau_p$ & 1 $\mu$s\\
\hline PHY header & 16 bytes & Slot time & 20 $\mu$s\\
\hline Payload size, E[PL] & 1024 bytes &SIFS & 10 $\mu$s\\
\hline ACK & 14 bytes&DIFS & 50 $\mu$s\\
\hline RTS & 20 bytes&EIFS & 300 $\mu$s\\
\hline CTS & 14 bytes & ACK,CTS timeout & 300 $\mu$s\\
\hline $m$ & 5 & $W$ & 32\\
\hline $T_s$ & 8.812 ms & $T_c$ & 8.812 ms\\
\hline\hline
\end{tabular}
 \label{tab.design.times}
\end{center}
\end{table}
This section focuses on simulation results for validating the
theoretical models and derivations presented in the previous sections. We have developed
a C++ simulator modelling both the DCF protocol details in 802.11b
and the backoff procedures of a specific number of independent
transmitting stations. The simulator also takes into account all real
operations of each transmitting station, including physical propagation
delays, etc.

Typical MAC layer parameters for the lowest rate IEEE802.11b are
given in Table~\ref{tab.design.times}~\cite{standard_DCF_MAC}. In
so far as the computation of the FER is concerned, it should be
noted that data transmission rate of various packet types differ.
For simplicity, we assume that data packets transmitted by
different stations are affected by the same FER.
%
%

The FER as a function of the SNR can be computed as follows:
\begin{equation}\label{fer_1}\small
P_e(SNR)=1-\left[1-P_e(PLCP,SNR)\right]\cdot
\left[1-P_e(DATA,SNR)\right]
\end{equation}
where,
\begin{equation}\label{fer_2}\small
P_e(PLCP,SNR)=1-\left[1-P_b(BPSK,SNR)\right]^{8\times PLCP},
\end{equation}
and
\begin{equation}\label{fer_3}\small
P_e(DATA,SNR)=1-\left[1-P_b(TYPE,SNR)\right]^{8\times (DATA+MAC)}.
\end{equation}
$P_b(BPSK,SNR)$ is the BER as a function of SNR for the lowest
data transmit rate employing DBPSK modulation, DATA denotes the
data block size in bytes, and any other constant byte size in
above expression represents overhead. Note that the FER,
$P_e(SNR)$, implicitly depends on the modulation format used.
Hence, for each supported rate, one curve for $P_e(SNR)$ as a
function of SNR can be generated. $P_b(TYPE,SNR)$ is modulation
dependent whereby the parameter $TYPE$ can be any of the following
$TYPE\in \{DBPSK,DQPSK,CCK5.5,CCK11\}$\footnote{The acronyms are
short for Differential Binary Phase Shift Keying, Differential
Quadrature Phase Shift Keying and Complementary Code Keying,
respectively.}.

For DBPSK and DQPSK modulation formats, $P_b(TYPE,SNR)$ can be
well approximated by \cite{SimonAlouini}:
\begin{equation}\small
\label{ber_function}
\frac{2}{\max(\log_{2}M,2)}\sum_{i=1}^{\max(\frac{M}{4},1)}\frac{1}{\pi}\int_{0}^{\frac{\pi}{2}}
\frac{1}{1+\gamma\frac{1}{\sin^2\theta}\log_{2}M\sin^2\left(\frac{(2i-1)\pi}{M}\right)}d\theta
\end{equation}
whereby $M$ is the number of bits per modulated symbols, $\gamma$
is the signal-to-noise ratio, and $\theta$ is the signal direction
over the Rayleigh fading channel.
\figura{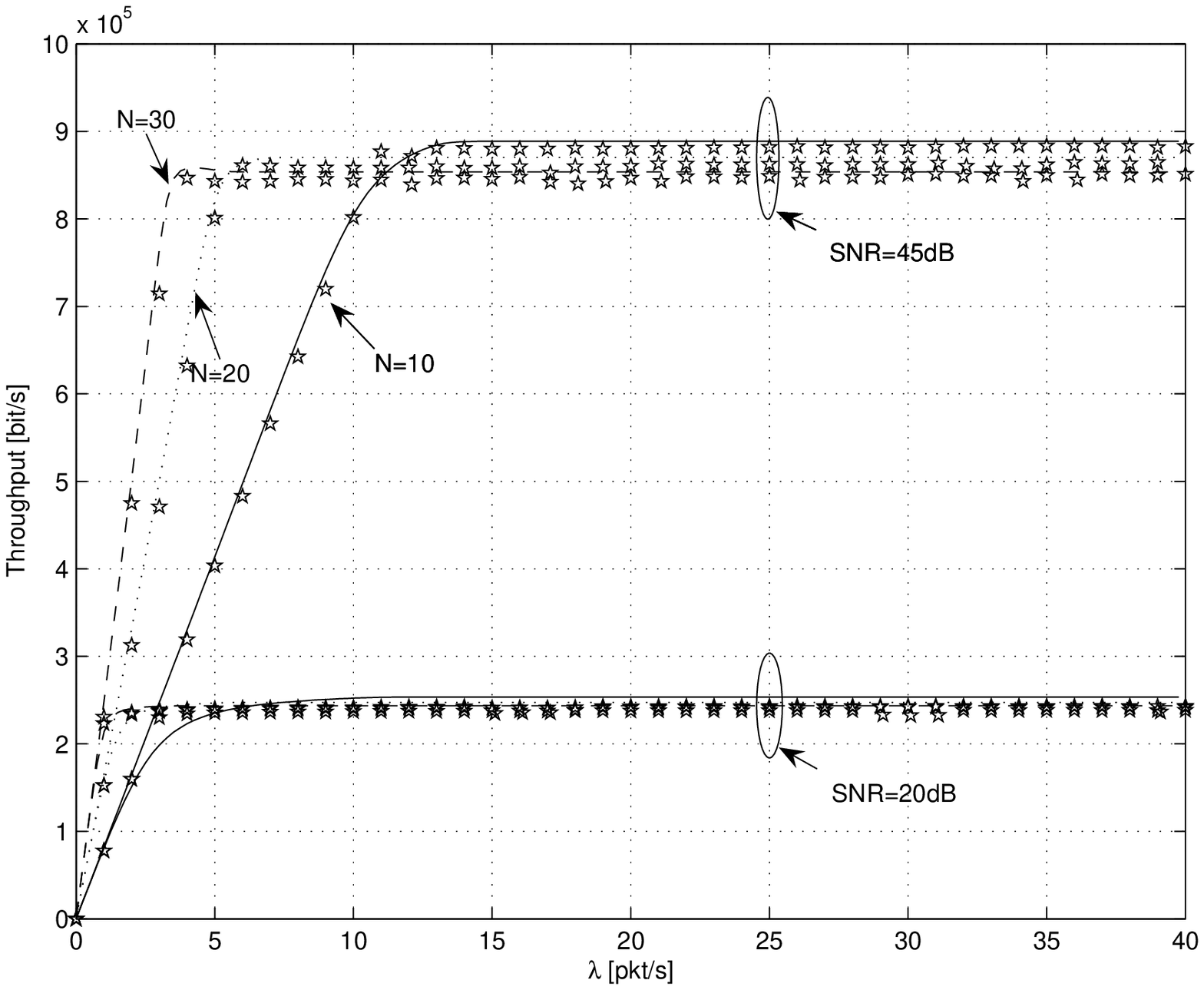}{Theoretical and simulated throughput for the
2-way mechanism as a function of the packet rate $\lambda$, for
three different number of contending stations and two different
values of SNR as noted in the legend. Capture thresholds is
$z_0=6$dB. Simulated points are identified by star-markers over
the respective theoretical curves. Payload size is 1024
bytes.}{Throgh_lambda_par_z0_6dB}

Fig.~\ref{Throgh_lambda_par_z0_6dB} shows the behavior of the
throughput as a function of $\lambda$, i.e., the packet rate, for
three different values of the number of contending stations and
for two values of SNR. The capture threshold is $z_0=6$dB. Beside
noting the throughput improvement achievable for high SNR, notice
that for a specified number of contending stations, the throughput
manifests a linear behaviour for low values of packet rates with a
slope depending mainly on the number of stations $N$. However, for
increasing values of $\lambda$, the saturation behavior occurs
quite fast. Notice that, as exemplified in~(\ref{eq:q_prob2}),
$q\rightarrow 1$ as $\lambda\rightarrow\infty$. Actually,
saturated traffic conditions are achieved quite fast for values of
$\lambda$ on the order of ten packets per second with a number of
contending stations greater than or equal to 10.

Fig.~\ref{Throgh_lambda_par_z0_24dB} shows the behavior of the
throughput as a function of $\lambda$, for three different values
of the number of contending stations and for two different SNRs. The capture threshold is
$z_0=24$dB. We can draw conclusions similar to the ones derived
for Fig.~\ref{Throgh_lambda_par_z0_6dB}. Upon comparing the curves
shown in Fig.s~\ref{Throgh_lambda_par_z0_6dB}
and~\ref{Throgh_lambda_par_z0_24dB}, it is easily seen that
capture effects allow the system throughput to be almost the same
independently from the number of stations in saturated conditions,
i.e., for high values of $\lambda$.

Fig.~\ref{Throgh_lambda_par_z0_24dB} also shows the presence of a
peak in the throughput as a function of $\lambda$, which
characterizes the transition between the linear and saturated
throughput. Such a peak tends to manifest itself for increasing
values of $\lambda$ as the number of stations $N$ decreases. A
comparative analysis of the curves shown in
Fig.s~\ref{Throgh_lambda_par_z0_6dB}
and~\ref{Throgh_lambda_par_z0_24dB} reveals that the peak of the
throughput tends to disappear because of the presence of capture
effects during transmission.
\figura{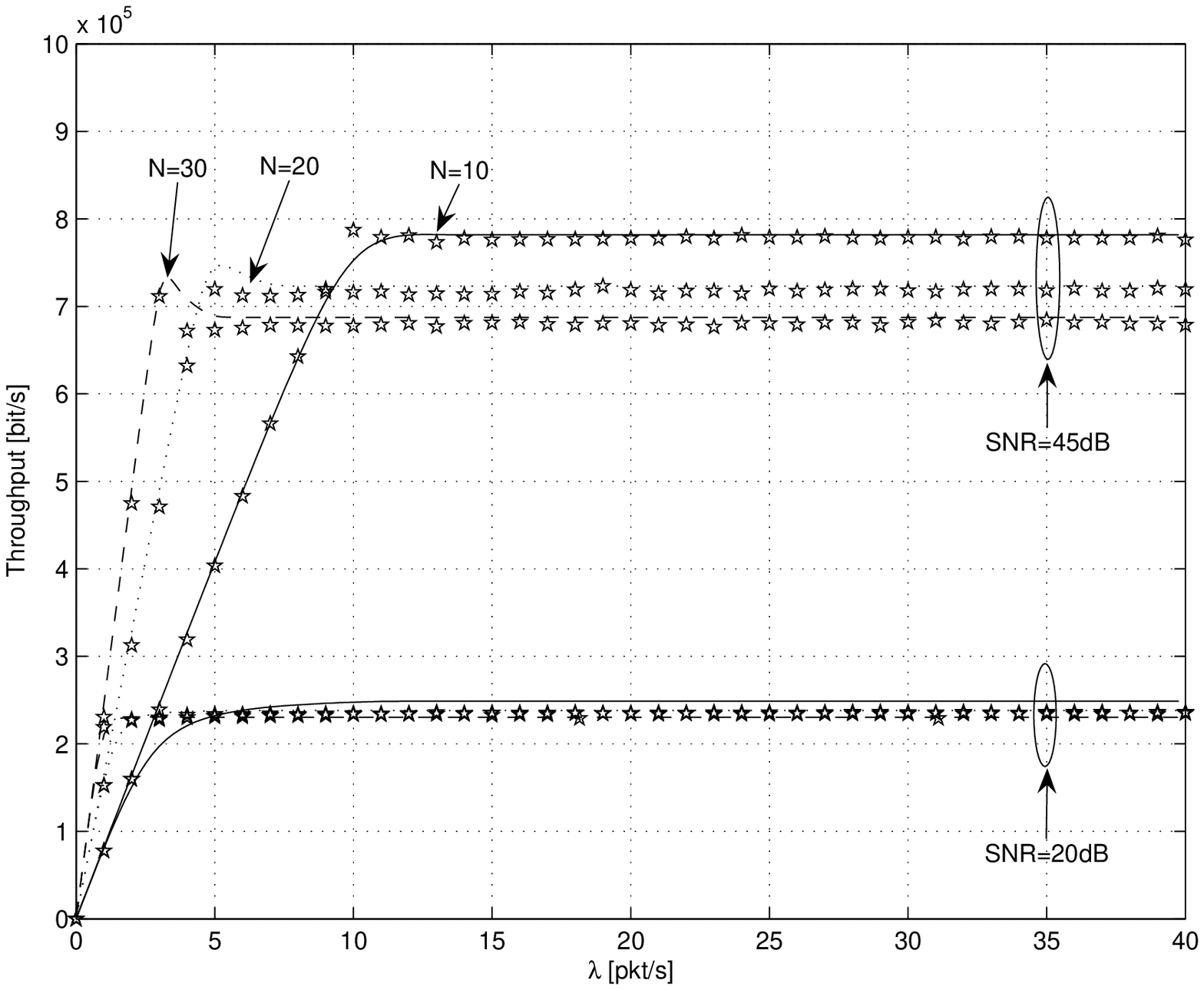}{Theoretical and simulated throughput for the
2-way mechanism as a function of the packet rate $\lambda$, for
three different number of contending stations and two different
values of SNR as noted in the legend. Capture threshold is
$z_0=24$dB. Simulated points are identified by star-markers over
the respective theoretical curves. Payload size is 1024
bytes.}{Throgh_lambda_par_z0_24dB}
\section{Conclusions}
In this paper, we have provided an extension of the Markov model
characterizing the DCF behavior at the MAC layer of the IEEE802.11
series of standards by accounting for channel induced errors and
capture effects typical of fading environments under unsaturated
traffic conditions. The modelling allows taking into consideration
the impact of channel contention in throughput analysis which is
often not considered or it is considered in a static mode by using
a mean contention period. Simulation results confirm the validity
of the proposed theoretical models.
\end{document}